\documentclass[conference]{IEEEtran}
\IEEEoverridecommandlockouts
\usepackage{cite}
\usepackage{amsmath,amssymb,amsfonts}
\usepackage{algorithmic}
\usepackage{graphicx}
\usepackage{textcomp}
\usepackage{xcolor}
\usepackage{url}
\usepackage{booktabs}
\usepackage{tabularx}
\usepackage{hyperref}
\usepackage{tikz}
\newcommand\copyrighttext{%
  \footnotesize © 2026 IEEE. Personal use of this material is permitted. Permission from IEEE must be obtained for all other uses, in any current or future media, including reprinting/republishing this material for advertising or promotional purposes, creating new collective works, for resale or redistribution to servers or lists, or reuse of any copyrighted component of this work in other works.}
\newcommand\copyrightnotice{%
\begin{tikzpicture}[remember picture,overlay]
  \node[anchor=south,yshift=10pt] at (current page.south) {\fbox{\parbox{\dimexpr\textwidth-\fboxsep-\fboxrule\relax}{\copyrighttext}}};
\end{tikzpicture}%
}
\def\BibTeX{{\rm B\kern-.05em{\sc i\kern-.025em b}\kern-.08em
    T\kern-.1667em\lower.7ex\hbox{E}\kern-.125emX}}

\begin{document}
\bstctlcite{IEEEexample:BSTcontrol}

\title{Plug-and-Play Sensor Evolution in Software-Defined Vehicles through Asset Administration Shells \\
\thanks{The authors would like to thank the German Federal Ministry of Research, Technology and Space (BMFTR) (under Grant Number: 16MEE0472) and the Chips Joint Undertaking for the financial support under Grant Agreement No: 101139789 (HAL4SDV). The responsibility for the content of this publication lies with the authors. 

Code can be found at: \url{https://github.com/akshaynarla/aas-pnp}}
}

\makeatletter
\newcommand{\linebreakand}{%
  \end{@IEEEauthorhalign}
  \hfill\mbox{}\par
  \mbox{}\hfill\begin{@IEEEauthorhalign}
}
\makeatother
\author{
\IEEEauthorblockN{{Akshay Narla, Nasser Jazdi and Michael Weyrich}}
\IEEEauthorblockA{
\textit{Institute of Industrial Automation and Software Engineering (IAS)} \\
{University of Stuttgart} \\
{Pfaffenwaldring 47, 70550 Stuttgart, Germany} \\
{E-Mail: \{akshay.narla, nasser.jazdi, michael.weyrich\}@ias.uni-stuttgart.de}}
}
\maketitle
\copyrightnotice
\begin{abstract}
Software-defined vehicles (SDVs) promise continuous software evolution, but sensor integration in deployed vehicles still depends on manual configuration, hardware-specific drivers, and repeated application-side adaptation. Existing standards address only parts of this problem. AUTOSAR standardizes selected sensor interfaces, the Vehicle Signal Specification (VSS) standardizes signal semantics, and Asset Administration Shells (AAS) standardize digital asset descriptions. However, these building blocks are rarely combined into an end-to-end onboarding path for sensors attached after vehicle deployment.

This paper presents an AAS-driven plug-and-play sensor integration framework for safety-supervised sensors. Each sensor is represented by an AAS that combines hardware identification, interface metadata, VSS mappings, and references to containerized driver and optional application software. After physical connection, an edge gateway identifies the device, retrieves the matching AAS, deploys the referenced software, and publishes the resulting data under stable vehicle signal paths. A prototype on a resource-constrained edge gateway connected to an RR100 UGV demonstrates automated integration of a USB camera and an ultrasonic sensor. Across repeated integration cycles, the prototype reaches operational state in about 2~s, shows no failed events, and keeps applications independent of hardware-specific interfaces. The results indicate that the introduction of AAS as an executable onboarding contract enables automated runtime deployment of sensor drivers in SDV environments.
\end{abstract}


\section{Introduction}

Software-defined vehicles (SDVs) decouple software from hardware to support over-the-air updates, continuous feature delivery, and hardware evolution across the vehicle lifetime. A key requirement of this transition is that applications remain independent of the specific sensing hardware installed in a given vehicle \cite{stuempfle2025, teixeira2024, narla2025hal}. Middleware surveys confirm the importance of modularity and service orientation, but also highlight unresolved challenges in lifecycle management, hardware abstraction, and integration of heterogeneous subsystems\cite{kluner2024}. 

These challenges become more visible after vehicle deployment if a sensor must be added or replaced without rebuilding the software stack. Existing standards address only parts of this problem. AUTOSAR Adaptive standardizes sensor interfaces and integration concepts for modern vehicle software, but remains largely design-time and tool-driven \cite{autosar_sensor_2024,autosar_rs_adi_2024}. VSS and Eclipse Kuksa provide a semantic vehicle signal model and a practical in-vehicle data layer, but assume that the sensor-side integration has already been completed \cite{covesa_vss_doc_2025, kuksa_userguide_2025}.

In parallel, the Asset Administration Shell (AAS) has matured into a standardized representation for digital assets and their interfaces in Industry 4.0 \cite{idta_part1_2025,idta_part2_2025}. Recent work on AAS discovery, interoperability, and engineering integration makes it a strong candidate for sensor self-description in post-deployment integration scenarios \cite{gordt2024discovery,zhang2025,gruener2023,kristan2025}.

This paper addresses this integration gap by providing an AAS-driven onboarding method and framework for non-safety-critical SDV sensors. We take AAS as an executable integration contract rather than a passive metadata container and combine it with VSS-based signal abstraction on a containerized vehicular edge gateway. The main contributions are:

\begin{itemize}
    \item Definition of an AAS model for deployable vehicular sensors that combines hardware identification, interface metadata, VSS mapping, and references to containerized software artifacts.
    \item Design of a fully automated integration workflow that turns a newly attached sensor into an operational VSS-compliant data source.
    \item Realization of a prototype on a resource-constrained vehicular edge gateway integrated with our UGV RR100.
    \item Evaluation on a real platform with heterogeneous sensors showing fast onboarding, repeated successful operation, and application-independent sensor replacement within the non-safety-critical scope of the prototype.
\end{itemize}

The remainder of this paper is structured as follows. Section~\ref{sec:related} reviews related work. Section~\ref{sec:requirements} defines the problem scope and requirements. Section~\ref{sec:approach} presents the onboarding approach. Section~\ref{sec:implementation} describes the prototype implementation. Section~\ref{sec:evaluation} reports the evaluation results. Section~\ref{sec:conclusion} concludes the paper.

\section{Related Work}\label{sec:related}
The literature relevant to this work can be organized into three groups: SDV data and interface layers, AAS as digital asset infrastructure, and plug-and-play integration in adjacent domains.

\subsection{Vehicular Data and Interface Layers}

Recent SDV research consistently treats the decoupling of software from hardware as a central architectural goal, with middleware, service orientation, and reusable communication layers as key enablers \cite{stuempfle2025,teixeira2024,kluner2024}. In parallel, AUTOSAR Adaptive standardizes sensor-related interfaces and requirements for automated driving software \cite{autosar_sensor_2024,autosar_rs_adi_2024}, while VSS and Kuksa Databroker provide a shared semantic model and an in-vehicle data layer for application access to vehicle signals \cite{covesa_vss_doc_2025,kuksa_userguide_2025}. Together, these works establish important foundations for interoperability and portability. However, they largely assume that the relevant driver or data provider has already been integrated. They do not address the onboarding of a previously unknown sensor that is newly connected to a deployed vehicle.

\subsection{AAS as Digital Asset Infrastructure}

The AAS has matured into a structured, machine-readable digital twin standard with stable metamodel and API specifications \cite{idta_part1_2025,idta_part2_2025}. Recent work has also emphasized discovery and lifecycle-aware integration as core concerns in AAS ecosystems \cite{gordt2024discovery,gruener2023}. Other studies show that AAS can be linked to IoT backends, engineering data systems, and broader digital twin toolchains, which strengthens its relevance as a runtime integration mechanism rather than a static documentation format \cite{zhang2025,kristan2025,quadrini2023}.

In automotive contexts, Schnauffer et al. demonstrate an AAS representation for the wiring harness system, showing that AAS can be applied to complex vehicle-related assets \cite{schnauffer2023}. More broadly, recent review work on digital twins and cyber-physical systems in commercial vehicles highlights unresolved interoperability and asset representation challenges, and identifies AAS as a promising mechanism to address them \cite{oliveira2024dtcps}. Similarly, our earlier research establishes the relevance of AAS for SDV interoperability and compares hardware abstraction layer concepts in SDV architectures \cite{narla2025interoperability,narla2025hal}. However, these works remain at the level of architecture, interoperability, and framework design. They do not demonstrate a complete in-vehicle onboarding that starts with physical sensor attachment and continues through to live signal publication.

\subsection{Plug-and-Play Integration in Adjacent Domains}

Plug-and-produce research in industrial automation addresses problems that are structurally similar to sensor onboarding in SDVs. In their review, Khabbazi et~al.\ identify interoperability, engineering effort reduction, and automatic configuration as persistent core goals~\cite{khabbazi2024}. Oliveira et~al.\ demonstrate practical plug-and-play transducer integration based on IEEE~1451 and IEC~61499~\cite{oliveira2022}. Pinheiro et~al.\ further show IEEE~1451-based self-identification and flexible edge-side integration for connected devices~\cite{pinheiro2022}. Together, these works show that machine-readable device descriptions and standardized integration workflows can reduce engineering effort in heterogeneous systems. 

However, the reviewed literature does not provide an end-to-end and automated integration pipeline for vehicle sensors. In particular, prior work does not show how a newly attached sensor can be identified, linked to deployable software, and exposed as a semantically standardized and application-ready data source within the SDV stack. This paper addresses that gap.

\section{Problem Definition and Requirements}
\label{sec:requirements}
This paper targets post-deployment integration of non-safety-critical or safety-supervised sensors for parking assistance, inspection cameras, environmental monitoring, or fleet-specific add-ons. It excludes hard real-time paths such as braking or primary perception, where design-time integration remains appropriate. Based on this and the gaps identified in Section II, we derive five onboarding requirements:
\begin{itemize}
    \item \textbf{R1 - Unambiguous sensor identification:} The platform must detect a newly attached sensor and associate it with a stable identifier without manual operator input.
    \item \textbf{R2 - Machine-readable self-description:} The platform must obtain a digital description that captures sensor identity, technical metadata, interfaces, and references to the software required for operation.
    \item \textbf{R3 - Deployable software binding:} The digital description must support the retrieval of the software components required to operate the sensor on the target platform.
    \item \textbf{R4 - Semantic normalization:} Sensor outputs must be mapped to stable vehicle-level signal paths so that they can be consumed consistently across applications.
    \item \textbf{R5 - Application-transparent runtime activation:} A newly attached sensor must become operational without rebuilding the vehicle software stack, and applications must consume normalized signals without depending on vendor-specific payloads or interfaces.
\end{itemize}

AAS directly addresses the machine-readable description and software binding requirements, while VSS and Kuksa provide the normalized vehicle-level interface required for application-transparent consumption. The main contribution of this work is the orchestration logic that turns these building blocks into a unified and automated integration workflow.

\begin{figure*}[!t]
    \centering
    \includegraphics[width=0.72\textwidth]{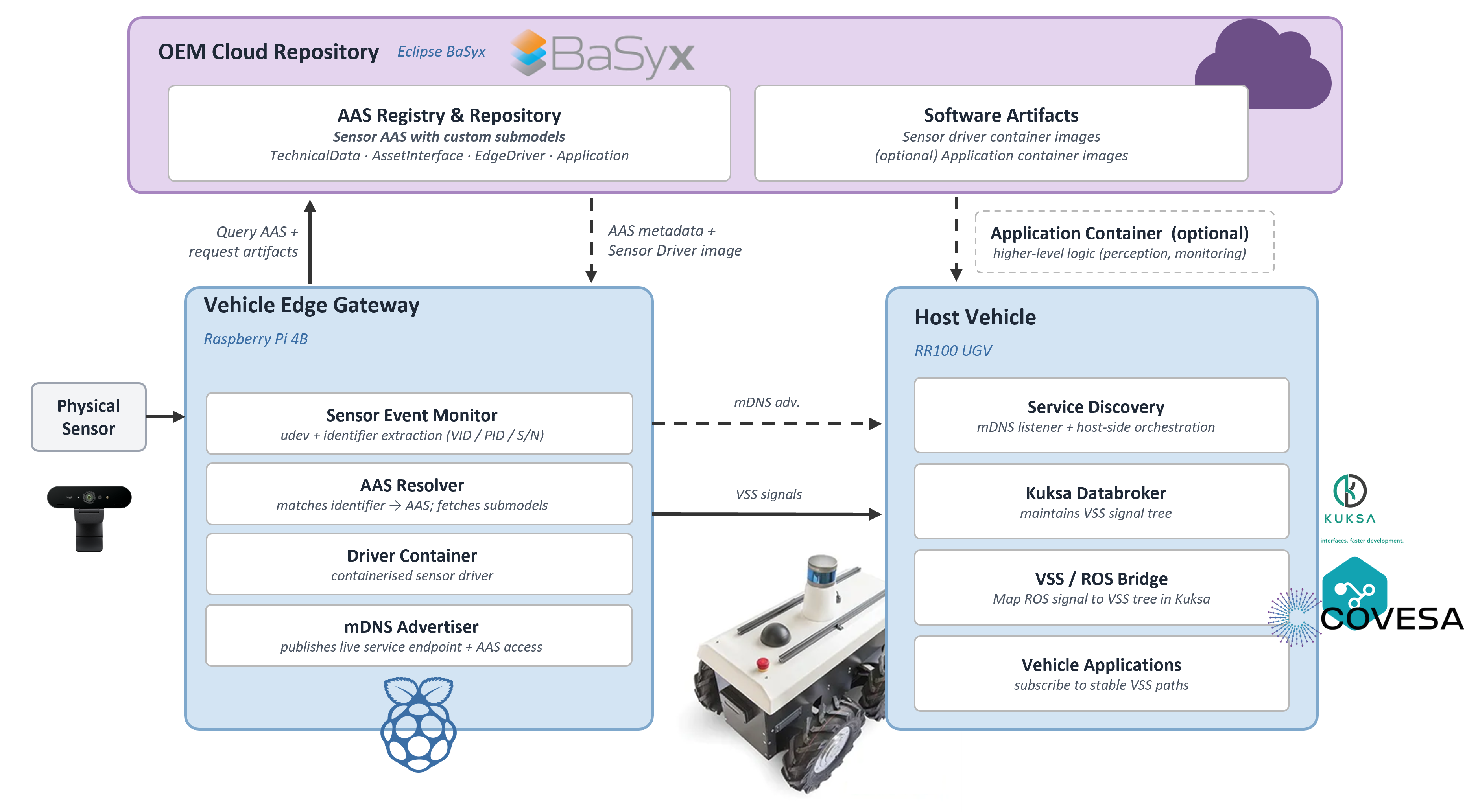}
    \caption{Prototype architecture showing the Original Equipment Manufacturer (OEM) Cloud Repository with AAS server, the Edge Gateway using Raspberry Pi, and the Host Vehicle Data Layer. The Edge Gateway detects hardware events, retrieves the matching AAS and driver image from the OEM cloud repository, advertises the resulting live AAS endpoint through mDNS, and publishes normalized VSS data toward the Host Vehicle.}
    \label{fig:architecture}
    \vspace{-4mm}
\end{figure*}

\section{AAS-Driven Integration Approach}\label{sec:approach}
The proposed approach is structured as an automated sensor onboarding workflow that separates sensor description, software provisioning, in-vehicle activation, and application-side signal consumption.

\subsection{Architecture Overview}

The architecture is organized into three domains, as illustrated in Fig.~\ref{fig:architecture}. It consists of an Original Equipment Manufacturer (OEM) Cloud Repository, a Vehicle Edge Gateway, and the Host Vehicle. This separation places sensor-specific integration logic close to the hardware while preserving a stable semantic interface toward the rest of the vehicle software.

The \textit{OEM Cloud Repository} provides the offboard infrastructure for AAS files and software artifacts. It stores the AAS entries used for sensor resolution together with the referenced driver and application images. This keeps sensor descriptions and deployable software under centralized OEM control.

The \textit{Vehicle Edge Gateway} realizes the onboarding logic. It detects newly attached sensors, extracts device identifiers, resolves the matching AAS entry, and activates the referenced software components. Once the driver stack is running, the gateway exposes a live AAS endpoint and advertises the service through mDNS so that it becomes discoverable without manual network-side configuration.

The \textit{Host Vehicle} provides the application-facing data layer. Kuksa Databroker exposes normalized vehicle signals through stable VSS paths, while the host-side orchestration logic discovers newly advertised services through mDNS and retrieves the configuration required to bind them into the existing vehicle software stack. This keeps application access stable even when the underlying sensor hardware changes.

The architecture described above defines the execution domains of the onboarding process. The following subsection specifies the AAS model used to represent sensor identity, deployment information, and signal integration.

\subsection{AAS Model for Deployable Sensors}

Each sensor is represented by an AAS containing four custom submodels. These submodels are not part of a standard AAS template. Instead, they were introduced specifically for the proposed onboarding process and to capture the necessary information required.

\begin{figure*}[t]
\centering
\includegraphics[
    width=0.8\textwidth,
]{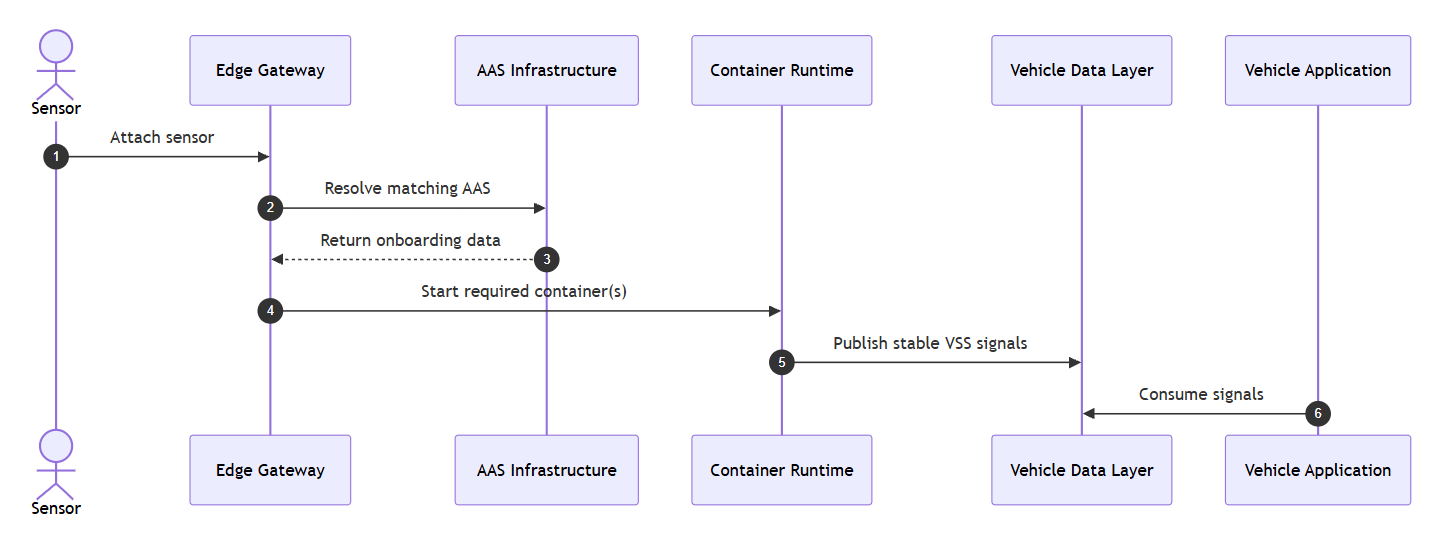}
\caption{Plug-and-Play Sensor Integration Workflow}
\label{fig:workflow}
\end{figure*}

\textit{TechnicalData} stores vendor ID, product ID, serial number, and sensor type. These fields enable deterministic matching between a physically attached device and its AAS entry through a constructed Global Asset ID.

\textit{AssetInterface} stores the runtime communication metadata required to expose the sensor at the vehicle level, including endpoint information, service metadata used for mDNS-based discovery, and the target VSS path. This submodel forms the bridge between the AAS-based sensor description and the vehicle signal model.

\textit{EdgeDriver} stores the information required to activate the sensor driver on the edge gateway, including the container image reference and the startup parameters needed at runtime.

\textit{Application} stores optional higher-level software artifacts that can be started after the driver becomes active.

The key design decision is to use AAS not only as descriptive metadata, but also as a deployable software contract for integration. In particular, the \textit{EdgeDriver} and \textit{Application} submodels extend the AAS from asset description toward executable orchestration intent, while \textit{AssetInterface} connects the resulting data stream to the VSS-based vehicle interface.

\subsection{Integration Workflow}
Fig.~\ref{fig:workflow} summarizes the sensor integration workflow. The novel contribution lies in orchestrating them into a single automated path from physical sensor attachment to application-ready signal availability.

The workflow starts with device detection on the edge gateway. When a new sensor is attached, the gateway reads the hardware descriptors that identify it and derives the corresponding asset identity. It then resolves the matching AAS entry and retrieves the software references and interface mappings defined in the relevant submodels.

The driver container is activated first. It communicates with the physical device, converts raw values into engineering-level data, and publishes that data under the VSS-aligned paths specified by the AAS. If the AAS also contains a reference to an application container, that component is activated afterwards. This second stage can encapsulate higher-level logic such as perception, monitoring, or event detection.

After startup, the edge gateway publishes the live service endpoint and AAS access point through mDNS. The host vehicle discovers this advertisement at runtime, retrieves the required configuration, and binds the new source into the vehicle data layer without manual endpoint setup. The final observable effect for applications is unchanged signal consumption through stable VSS paths. The resulting workflow enables both functional interoperability across sensor variants and deployment flexibility for new sensor types, without changes to the application layer.

\section{Prototype Implementation}\label{sec:implementation}
The previous sections described the onboarding concept at the architectural and workflow levels. The following section summarizes its realization as a prototype used for evaluation.

\subsection{Platform and Sensors}

The prototype was implemented on a resource-constrained edge gateway\footnote{Raspberry Pi~4B (4GB RAM, Cortex-A72, Gigabit Ethernet). Selected for reproducibility. The framework imposes no platform-specific dependencies.} connected via Gigabit Ethernet to a ROS-based RR100 unmanned ground vehicle. The RR100 provides front-facing sensing but no rear-mounted sensing, which creates a realistic use case for plug-and-play integration of reversing-assistance sensors.

Two heterogeneous USB-connected sensors were used: a Logitech C270 camera, representative of a high-bandwidth sensing path, and an ultrasonic distance sensor connected through an RS-485-to-USB adapter, representative of a scalar measurement path. Together, they evaluate the onboarding architecture under two different runtime profiles, namely stream-oriented sensing with downstream processing and direct publication of low-bandwidth scalar measurements.

\subsection{Software Stack}

The prototype uses replaceable open-source tools as its reference realization.
Eclipse BaSyx \cite{basyx_project_2025} hosts the AAS repository and serves AAS content over a REST API. On the edge gateway, \texttt{udev} detects hot-plug events, an HTTP client communicates with the AAS backend, Docker manages the lifecycle of referenced software components, and an mDNS advertiser exposes newly activated services at runtime. On the vehicle side, Kuksa Databroker maintains the VSS signal tree, while a VSS--ROS bridge and mDNS-based service discovery integrate the new source into the host software stack.

An automated AAS generation pipeline, based on \cite{yuchen_aas}, constructs sensor-specific AAS files from device metadata and populates the four custom submodels described above. The generated AAS instances were validated against the AAS metamodel and checked using aas-test-engines \cite{aas_test}.

\subsection{Containerized Sensor Drivers}

For the ultrasonic sensor, the driver container reads serial measurements, converts the raw values into engineering units, and publishes distance signals to the VSS path defined in the AAS. An associated application container implements a simple three-zone parking-assistance logic.

For the camera, the driver container captures image frames and exposes the stream to a second container that performs higher-level processing. In the prototype, this application layer carries out real-time object detection and publishes semantically meaningful safety information to the VSS tree. Together, the two cases show that the architecture can onboard both direct signal sources and sensor-dependent processing chains.

Multi-stage Docker builds reduce image size, and device passthrough grants containers access only to the required hardware interface.

\subsection{Repository Trust and Execution Isolation}

The prototype assumes an OEM-controlled repository as the source of plug-and-play metadata, but still separates retrieval from execution. AAS retrieval is mediated through a reverse proxy, while each sensor driver runs in its own container with access limited to the required device node. This keeps orchestration logic independent of sensor-specific code and reduces the impact of driver-level faults.

This design does not by itself solve the broader security problem of runtime deployment. In a production setting, authenticated AAS retrieval, signed container images, policy checks, and other safety-related considerations before startup would be required.

\section{Evaluation}\label{sec:evaluation}

This section evaluates the proposed onboarding approach and R1-R5 through four questions: 
(1) how quickly device identification (R1), AAS resolution (R2), container activation (R3), and VSS publication (R4) make a new sensor operational, 
(2) if the workflow remains robust across repeated insertions, 
(3) if application transparency (R5) is preserved when the underlying sensor changes, and 
(4) what compute overhead is introduced on the edge gateway. Twenty integration cycles were executed across camera and ultrasonic connections. Each cycle started with physical connection and ended when sensor data became available in the vehicle data layer.

\begin{figure}[!t]
    \centering
    \includegraphics[width=0.85\columnwidth]{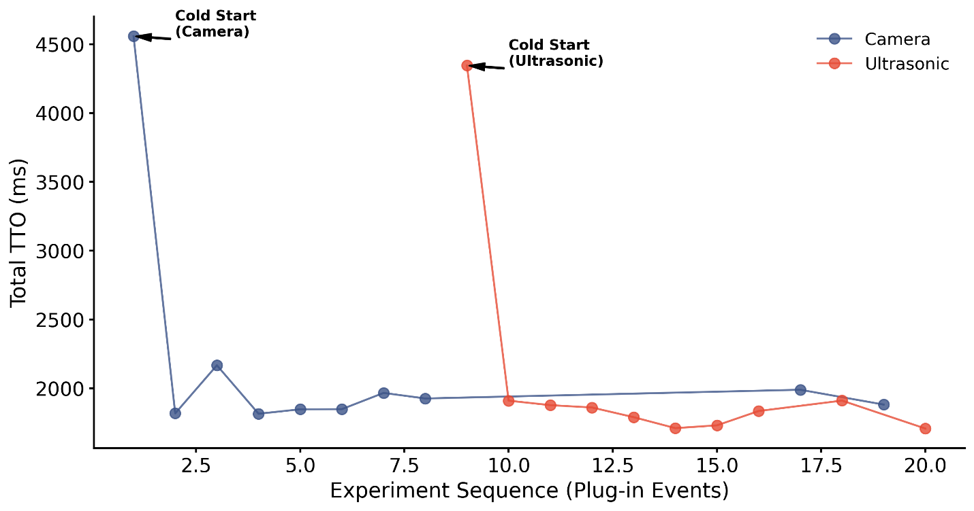}
    \caption{Time to operation across 20 consecutive integration cycles.  First-time onboarding includes image-retrieval cold starts, while subsequent re-insertions converge to a stable operating range near 2\,s.}
    \label{fig:tto}
    \vspace{-4mm}
\end{figure}

\subsection{Integration Latency}

Fig.~\ref{fig:tto} reports first-use image-retrieval cold starts separately; after image caching, the mean time to operation was $2.1 \pm 0.2$\,s for the ultrasonic sensor and $2.2 \pm 0.3$\,s for the camera in steady state. A phase-level inspection showed that AAS retrieval remained below 120\,ms, with a mean of 97\,ms. The dominant latency contribution came from container startup and application initialization rather than from digital asset lookup. This indicates that the AAS-based discovery stage does not form the critical path of activation and that further reductions can be pursued through image prefetching or local caching.

\subsection{Repeated Operation and Robustness}

Across all 20 repeated cycles, no event failed. Every cycle completed device identification, AAS retrieval, container launch, mDNS advertisement, and VSS publication automatically. Hardware recognition was further checked with five USB devices: two cameras of the same model, one camera of a different model, and two ultrasonic sensors.

Fig.~\ref{fig:compatibility} summarizes the device-compatibility checks. Cameras of the same model were matched successfully to the same AAS template, the unregistered camera was rejected correctly, and both ultrasonic sensors were classified through their shared RS-485 interface. This is an important result that shows the framework not only onboards compatible sensors automatically, but it also avoids false-positive activation for incompatible devices.

\begin{figure}[!t]
    \centering
    \includegraphics[width=0.83\columnwidth]{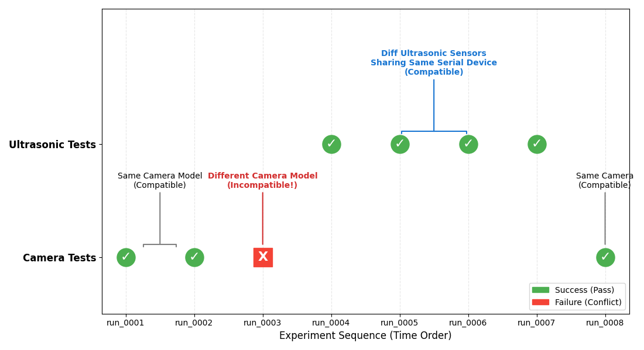}
    \caption{Compatibility checks across representative camera and ultrasonic onboarding runs. The incompatible camera is rejected intentionally because no matching AAS entry exists.}
    \label{fig:compatibility}
    \vspace{-4mm}
\end{figure}

These results should be interpreted only within the scope of the prototype. The experiments were conducted in a controlled environment with stable networking and a reachable AAS repository, so they demonstrate functional robustness of the onboarding workflow rather than reliability.

\subsection{Application-Independent Sensor Replacement}

A central claim of the proposed approach is that applications consume semantic vehicle signals rather than hardware-specific interfaces. This property was examined with two reversing-assistance use cases on the RR100 platform. When the ultrasonic sensor is connected, its driver container starts and publishes distance measurements to a parking-assistance VSS path. The corresponding vehicle-side logic consumes this path without knowledge of the underlying serial interface.

When the USB camera is connected instead, the framework detects the new device, retrieves its AAS, and deploys the camera driver together with the higher-level processing container. Again, the consuming side interacts only with the resulting vehicle-level signals and does not depend on the device-specific transport or startup procedure. The prototype, therefore, preserves application independence across heterogeneous sensor modalities, even though the underlying processing differs.

\subsection{Operational Interpretation and Validity Limits}

The orchestration logic remained lightweight on the Raspberry Pi edge gateway. CPU overhead attributable to onboarding stayed below 3\%. This indicates that the additional orchestration, AAS lookup, and container startup logic are operationally feasible on a constrained edge node within the scope of the prototype.

The main threats to validity are the limited number of sensor types, the use of a single edge platform, and the local-network deployment setting. The study does not quantify behavior under intermittent repository access, concurrent multi-sensor onboarding, or stronger security controls such as signed image verification. It also does not quantify behavior under varied network conditions, although retrieval is already separated from execution for running sensors. These limitations define the current evidence boundary rather than a contradiction of the contribution.

Overall, the results demonstrate that AAS-based plug-and-play sensor integration is feasible on a resource-constrained vehicular edge platform, with sub-3-second activation, correct rejection of incompatible devices, and consistently successful operation under controlled conditions. The time-to-operation is dominated by container startup rather than AAS lookup or orchestration, which suggests that the most effective optimization targets are image pre-caching and reduced container cold-start overhead. From an embedded-systems perspective, the framework complements AUTOSAR and remains targeted at non-safety-critical post-deployment sensor additions. A summary of the result is in \ref{tab:results}.

\begin{table}[!t]
\caption{Summary of prototype evaluation results}
\label{tab:results}
\centering
\footnotesize
\renewcommand{\arraystretch}{1.12}
\begin{tabular}{ll}
\toprule
Metric & Result \\
\midrule
Integration cycles & 20 \\
Mean TTO (ultrasonic) & $2.1 \pm 0.2$\,s \\
Mean TTO (camera) & $2.2 \pm 0.3$\,s \\
AAS lookup latency & 97\,ms mean, $<120$\,ms max \\
Failed integration events & 0/20 \\
CPU overhead & $<3\%$ on the edge gateway \\
Sensor modalities & USB camera, ultrasonic sensor \\
\bottomrule
\end{tabular}
\vspace{-2mm}
\end{table}

\section{Conclusion and Future Work}\label{sec:conclusion}

This paper presented an AAS-driven plug-and-play sensor integration framework for software-defined vehicles. The approach combines AAS-based sensor description, software binding, mDNS-based service discovery, and VSS-based signal abstraction into a workflow that turns a newly attached sensor into an application-ready vehicle data source. To support this workflow, the paper introduced a custom AAS model for deployable sensors together with an automated integration process spanning sensor identification, software activation, runtime discovery, and normalized signal publication.

The prototype on a resource-constrained edge gateway connected to our RR100 UGV demonstrated feasibility within the targeted non-safety-critical scope. With respect to the requirements defined in Section~\ref{sec:requirements}, the prototype confirmed unambiguous identification of attached devices (R1), machine-readable resolution through AAS (R2), automated retrieval and deployment of referenced containers (R3), semantic normalization via stable VSS paths (R4), and application-transparent activation without stack rebuilding (R5). Across repeated cycles, the prototype reached operational state in about 2~s and showed consistently successful operation. The main contribution is an AAS-based onboarding method and framework that treats AAS as an executable integration contract for post-deployment sensor evolution.

Future work will focus on three directions. First, the framework should be evaluated under more demanding safety-relevant scenarios, including intermittent repository access, sensor hot-swap, and unexpected disconnection. Second, security should be strengthened through authenticated AAS retrieval, signed container images, and policy checks before startup. Third, the approach should be extended to additional interface types and broader vehicular software environments, including closer integration with safety-relevant AUTOSAR Adaptive and other production-oriented SDV platforms.

\bibliographystyle{IEEEtran}
\bibliography{bib}

\end{document}